\documentclass[11pt,a4paper]{article}

\usepackage[T1]{fontenc}
\usepackage[utf8]{inputenc}
\usepackage[english]{babel}

\usepackage[a4paper,margin=1in]{geometry}
\usepackage{setspace}
\usepackage{parskip}
\usepackage{ragged2e}
\usepackage{quantikz}
\usepackage{amsmath,amssymb,amsfonts,amsthm}
\usepackage{bm}
\usepackage{braket}

\usepackage{graphicx}
\usepackage{multirow}
\usepackage[title]{appendix}
\usepackage{xcolor}
\usepackage{textcomp}
\usepackage{booktabs}
\usepackage{algorithm}
\usepackage{algpseudocode}
\usepackage{listings}

\usepackage[numbers,sort&compress]{natbib}
\usepackage[colorlinks=true,linkcolor=blue,citecolor=blue,urlcolor=blue]{hyperref}
\usepackage[nameinlink,capitalise,noabbrev]{cleveref}
\usepackage{orcidlink}
\usepackage{authblk}

\theoremstyle{definition}

\theoremstyle{remark}

\title{\textbf{A Complex-Valued Continuous-Variable Quantum Approximation Optimization Algorithm (CCV-QAOA)}}

\author[1]{Raneem Madani\,\orcidlink{0009-0006-6757-0181}}
\author[1]{Abdel Lisser\,\orcidlink{0000-0003-1318-6679}}
\author[1]{Zeno Toffano\,\orcidlink{0000-0001-8594-3291}}

\affil[1]{CNRS, CentraleSup\'elec, Laboratoire des Signaux et Syst\`emes (L2S), Universit\'e Paris-Saclay, 91190 Gif-sur-Yvette, France}

\date{2026}

\begin{document}

\maketitle

\begin{center}
Raneem Madani: \texttt{raneem.madani@centralesupelec.fr} \\
Abdel Lisser: \texttt{abdel.lisser@centralesupelec.fr} \\
Zeno Toffano: \texttt{zeno.toffano@centralesupelec.fr}
\end{center}

\begin{abstract}
Continuous-variable (CV) quantum systems offer a natural framework for continuous optimization through their infinite-dimensional Hilbert spaces. In this paper, we propose the Complex Continuous-Variable Quantum Approximate Optimization Algorithm (CCV-QAOA), a variational framework operating in the complex domain that optimizes over complex decision variables. The method efficiently solves real and complex multivariate optimization problems. To demonstrate its versatility, we apply CCV-QAOA across a broad suite of optimization use cases, including convex quadratic minimization, scaling studies with circuit depth and cutoff dimension, constrained quadratic programs using penalty constructions, and non-convex benchmarks such as the Styblinski--Tang function and complex quartic landscapes.
\end{abstract}

\section{Introduction}\label{sec1}
In many applications, optimization problems are encountered in which the goal is to minimize or maximize an objective function subject to constraints. Over the past few decades, quantum computing has enabled the development of powerful algorithms for these problems. Much of the work has focused on cases with discrete decision variables, and more recently on continuous real-valued optimization. However, many real-world problems are naturally expressed with complex-valued decision variables, where both magnitude and phase play important roles. Examples arise in areas such as signal processing \cite{peng2024beamforming}, Mimo \cite{tassouli2023maximizing, adasme2019variable}, wierless \cite{adasme2023stochastic}, quantum information and communication \cite{shao2024solving, lauro2025optimization}, chemistry \cite{zheng2024unleashed}, and quantum control \cite{dong2010quantum}. In these cases, rewriting complex variables as pairs of real variables hides the mathematical and physical structure, and also adds extra computational cost.

On the classical computing side, a variety of optimization models involving complex decision variables have been investigated. Early work introduced deterministic formulations—complex linear, nonlinear, and quadratic programs—together with corresponding duality theories \cite{levinson1966linear, abrams1972nonlinear, craven1973duality, datta1984duality, ferrero1992nonlinear}. These were later complemented by approaches grounded in Cauchy-Riemann CR-calculus and unconstrained complex optimization \cite{kreutz2009complex, sorber2012unconstrained}. More recent developments have shifted toward stochastic models that incorporate uncertainty directly through probabilistic feasibility requirements, giving rise to complex chance-constrained programming (CCCP) \cite{madani2025chance}. Solving such problems typically relies on gradient-based schemes \cite{ancona2019gradient} or derivative-free methods. Among the latter, Covariance Matrix Adaptation Evolution Strategy (CMA-ES) has emerged as particularly effective for highly nonconvex or black-box formulations, where the objective function is not analytically available \cite{hansen2001completely, hansen2016cma}.

Quantum computing leverages superposition, interference, and entanglement to offer potential advantages in optimization, motivating variational algorithms such as the Variational Quantum Eigensolvers (VQE) \cite{tilly2022variational}. However, most existing methods target qubit architectures, where outputs are fundamentally discrete. As a result, continuous problems are typically reformulated as QUBO or polynomial binary programs and compiled into gate circuits \cite{li2008quantum,talbi2017new,wang2025direct, o2007optical}. While theoretically powerful, this binary encoding introduces deep circuits and high resource requirements on near-term devices \cite{moll2018quantum,preskill2018quantum}. The challenge becomes especially severe for continuous-variable tasks naturally defined over Hilbert spaces such as $\mathbb{R}^n$ and $\mathbb{C}^n$ \cite{stein2010complex}, where nonconvex problems are NP-hard \cite{murty1987np,ahmadi2013np}. These scalability limits motivate moving beyond qubit-based discretization toward continuous-variable quantum architectures.

The continuous-variable (CV) model offers an alternative to qubit-based computing by encoding information in quantum field states, such as the electromagnetic field, making it ideal for photonic hardware~\cite{braunstein2005quantum}. Its observables, position $\hat{x}$ and momentum $\hat{p}$, have continuous spectra linked to the wave nature of quantum systems, and qubit computation can be embedded within this framework without loss of power~\cite{weedbrook2012gaussian}. By representing information in the continuous quadratures of Bosonic modes, CV architectures operate in a naturally infinite-dimensional Hilbert space, allowing compact encodings and smooth optimization landscapes that are well suited for variational and machine-learning tasks. Recent studies have demonstrated fundamental optimization primitives in CV systems~\cite{schuld2019quantum,arrazola2019machine}, yet their full potential—especially for continuous and complex-valued optimization—remains largely unexplored. Extending the Quantum Approximate Optimization Algorithm (QAOA)~\cite{farhi2014quantum} to the CV domain provides a promising route for continuous optimization on near-term photonic quantum devices~\cite{verdon2019quantum,enomoto2023continuous}.

Quantum computing can be built on different hardware platforms, each with its own advantages and limitations. CV systems use bosonic modes, such as light or microwave fields, instead of two-level qubits. Optical photonic systems \cite{lee2005inspirations} work at room temperature and are stable against decoherence, making them ideal for communication, but their weak nonlinearities make universal operations difficult. Microwave cavity systems \cite{bradley2003microwave}, based on superconducting circuits, allow very precise control and error correction but require extremely low temperatures and cannot transmit signals over long distances. Trapped-ion systems mainly use qubits but can also store information in their continuous motional modes \cite{gan2020hybrid}; they have excellent coherence but are hard to scale. A newer platform, the microcomb-driven silicon photonic system \cite{shu2022microcomb}, combines integrated optical frequency combs with CMOS-compatible silicon photonics to achieve compact, low-power, and scalable CV photonic processing. It operates at room temperature and enables multi-channel optical and microwave signal control, though nonlinear gate implementation remains a challenge \cite{brady2024advances}.

In this paper, we present a novel variational quantum optimization framework that operates natively in the complex domain, the Complex Continuous-Variable Quantum Approximate Optimization Algorithm (CCV-QAOA). This framework generalizes the CV-QAOA~\cite{verdon2019quantum} algorithm to handle real-valued objective functions with complex-valued decision variables, enabling a richer and more expressive optimization landscape. To validate the proposed approach, we investigate representative complex optimization problems, both constrained and unconstrained, convex and non-convex. Constraints are reformulated into equivalent unconstrained forms to enable efficient implementation within the variational circuit, while universal gate sets are employed to approximate higher-order operators. Moreover, CCV-QAOA can solve multivariate real-valued optimization problems using only half the number of modes required in conventional CV formulations, achieving compact yet expressive encoding. Theoretical analysis and numerical experiments conducted within a finite Fock Hilbert space with dimention $D$, include unconstrained quadratic problems, linearly constrained systems, and non-convex benchmarks such as the Styblinski–Tang function, demonstrating that CCV-QAOA can reach the optimal solution without requiring an infinite-dimensional representation. These results confirm the algorithm’s ability to efficiently navigate complex energy landscapes through universal gate decompositions of the corresponding quartic Hamiltonians.

This paper is organized as follows. Section~\ref{sec2} introduces the complex-valued CV model for optimization problems: we first review the standard QAOA framework, then recall the fundamentals of CV quantum computation and gate universality, and finally establish the mapping between complex variables and phase-space dynamics that underpins CCV-QAOA. Section~\ref{sec3} presents the full formulation of the CCV-QAOA algorithm, including the variational ansatz, backend choices (Gaussian and Fock), measurement strategies, and classical optimization loop. Section~\ref{sec4} illustrates the method on several case studies, covering unconstrained and constrained quadratic programs, multivariate real-valued problems with reduced mode counts, and non-convex benchmarks such as the Styblinski–Tang function, and compares CCV-QAOA with CV-QAOA. Finally, Section~\ref{sec5} discusses the main findings, resource–accuracy trade-offs arising from cutoff and depth, and outlines limitations and directions for future work.
\section{Complex Valued CV-Model for Optimization Problems} \label{sec2}
In this section, we first present the general formulation of the Quantum Approximate Optimization Algorithm (QAOA) for optimization problems. We then review the fundamentals of CV quantum computation and discuss gate universality within this framework, highlighting how these gates can be integrated into the CV-QAOA structure. Finally, we introduce the model connecting complex-valued optimization to phase-space dynamics, establishing the theoretical foundation for extending QAOA to the complex continuous-variable domain (CCV-QAOA).
\subsection{The QAOA Algorithm}
The Quantum Approximate Optimization Algorithm (QAOA) is a variational quantum-classical framework designed to solve combinatorial \cite{farhi2014quantum} and continuous optimization problems \cite{verdon2019quantum}. Its core idea is to map a classical objective function $f(x)$ into a quantum cost Hamiltonian $\hat{H}_C$ whose ground state encodes the optimal solution. To explore the solution space, a complementary mixer Hamiltonian $\hat{H}_M$, chosen such that $[\hat{H}_M,\hat{H}_C]\neq 0$, is introduced. This non-commutativity ensures that the quantum dynamics are able to leave the initial configuration and explore alternative solutions.

A QAOA circuit of depth $q$ is constructed as an alternating sequence of unitaries generated by the cost and mixer Hamiltonians. Starting from an easily preparable initial state $\ket{\psi_0}$ (commonly the uniform superposition), the system evolves under successive applications of these unitaries, producing a parametrized quantum state  
\begin{align}
\ket{\psi(\boldsymbol{\gamma}, \boldsymbol{\beta})} 
= \prod_{j=1}^p e^{-i \beta_j \hat{H}_M} e^{-i \gamma_j \hat{H}_C} \ket{\psi_0}, \label{qaoa}
\end{align}
where $(\boldsymbol{\gamma}, \boldsymbol{\beta})$ are variational parameters of the algorithm.

For a given choice of $(\boldsymbol{\gamma},\boldsymbol{\beta})$, the quality of the state is evaluated by measuring the cost Hamiltonian. In practice, the expectation value
\begin{align}
\mathcal{L}^{(t)} = \bra{\psi^{(t)}} \hat{H}_C \ket{\psi^{(t)}} \label{expectation}
\end{align}
is estimated from repeated measurements (shots) on the output state $\ket{\psi^{(t)}}$, yielding a sampled cost that is fed back to a classical optimizer. The optimizer then updates the parameters $(\boldsymbol{\gamma}, \boldsymbol{\beta})$ and the procedure state preparation, evolution under $\hat{H}_C$ and $\hat{H}_M$, measurement, and classical update are repeated until convergence to a locally optimal set of parameters.

As the depth $q$ increases, the algorithm becomes more expressive and can approximate the true ground state with higher fidelity, albeit at the cost of deeper circuits. This framework is not affected to the type of optimization variables, making it applicable to both discrete systems, implemented with qubits and discrete operators, and continuous systems, implemented with qumodes and continuous operators. The distinction lies in how the initial state, cost Hamiltonian, mixer dynamics, and variable encoding are formulated, as well as in the overall circuit structure. In the following section, we introduce the CV setting, a representation that provides a natural foundation for extending QAOA to complex-valued optimization.
\subsection{Universality in the CV Model}
In CV quantum computing, information is encoded in the states of bosonic modes often referred to as qumodes whose Hilbert spaces are infinite-dimensional \cite{lloyd1999quantum, braunstein2005quantum}. The most elementary CV system is the quantum harmonic oscillator, described by the canonical ladder operators $\hat{a}$ and $\hat{a}^\dagger$. It is often convenient to represent CV quantum states using the quadrature operators $\hat{x}$ and $\hat{p}$, which satisfy the canonical commutation relation $[\hat{x}, \hat{p}] = i\hbar$ and serve as the quantum analogs of position and momentum \cite{nielsen2010quantum, weedbrook2012gaussian}. In the position representation, a quantum state is specified by a wave function $\psi(x)$, with $|\psi(x)|^2$ giving the probability density of measuring the system at position $x$. Equivalently, in the momentum representation, the state is described by $\phi(p)$, the Fourier transform of $\psi(x)$. More generally, $x$ and $p$ can be viewed as the real and imaginary components of a quantum field, such as a mode of the electromagnetic field.

The vacuum state $\ket{0}$ represents the ground state of the harmonic oscillator and serves as the reference state for CV quantum theory. In phase space, the vacuum exhibits a Gaussian Wigner distribution centered at the origin \cite{weedbrook2012gaussian}. More general states can be obtained by unitary evolution of the vacuum,
\begin{equation}
    \ket{\psi(t)} = e^{-i t H} \ket{0},
    \label{eq:evolution}
\end{equation}
where $H$ is a bosonic Hamiltonian expressed as a function of quadrature operators $\hat{x}_i$ and $\hat{p}_i$, and $t$ is the evolution time \cite{braunstein2005quantum}. 

As shown in \cite{lloyd1999quantum}, the universal gate set for CV quantum computation is ${D, R, S, BS, V}$, where $D$ (displacement), $R$ (rotation), $S$ (squeezing), and $BS$ (beamsplitter) are Gaussian gates, while $V$ (the cubic phase gate), or equivalently the Kerr interaction provides the essential non-Gaussian resource required for universality. Higher-order Hamiltonians can be systematically constructed from these elementary gates using concatenation and commutator identities. Table~\ref{tab:cv_gates_symbols} summarizes these fundamental operations: the first column lists the universal Hamiltonian generators, the second column shows the corresponding gate, the third column provides its circuit symbol, and the fourth column indicates whether the operation is Gaussian or non-Gaussian. Together, these gates form the core building blocks of CV quantum circuits.

\begin{table}[tb]
\centering
\caption{List of CV quantum gates with their unitary forms and circuit symbols}
\label{tab:cv_gates_symbols}
\begin{tabular}{|c|c|c|c|}
\hline
\textbf{Cost Hamiltonian $\hat{H}_c$} & \textbf{Gate} & \textbf{Symbol} & \textbf{Type} \\
\hline
$a\hat{x} + b\hat{p}$ & Displacement &
\begin{quantikz}
\lstick{} & \gate{D} & \qw
\end{quantikz} & Gaussian \\
\hline
$a(\hat{x}^2+\hat{p}^2)$ & Rotation &
\begin{quantikz}
\lstick{} & \gate{R} & \qw
\end{quantikz} & Gaussian \\
\hline
$a(\hat{x}\hat{p}+\hat{p}\hat{x})$ & Squeezing &
\begin{quantikz}
\lstick{} & \gate{S} & \qw
\end{quantikz} & Gaussian \\
\hline
$a \hat{x}_1 \hat{x}_2$ & Controlled-phase & 
\begin{quantikz}
\lstick{} & \ctrl{1} & \qw \\
\lstick{} & \gate{Z} & \qw
\end{quantikz} & Gaussian \\
\hline
$a \hat{x}_1 \hat{p}_2$ & Controlled-$X$ gate &
\begin{quantikz}
\lstick{} & \ctrl{1} & \qw \\
\lstick{} & \gate{X} & \qw
\end{quantikz} & Gaussian \\
\hline
$a\hat{x}^3$ & Cubic phase &
\begin{quantikz}
\lstick{} & \gate{V} & \qw
\end{quantikz} & Non-Gaussian \\
\hline
$a((\hat{x}-i\hat{p})(\hat{x}+i\hat{p}))^2$  & Kerr &
\begin{quantikz}
\lstick{} & \gate{K} & \qw
\end{quantikz} & Non-Gaussian \\
\hline
\end{tabular}
\end{table}

A simple geometric construction illustrates how new Hamiltonian transformations emerge from repeated applications of operations within a given set. Specifically, applying the Hamiltonians $\hat{A}$ and $\hat{B}$ in sequence for short time intervals $\delta t$ leads to an effective evolution generated by their commutator. For example, applying $\hat{A}$ for $\delta t$, then $\hat{B}$, followed by $-\hat{A}$ and $-\hat{B}$, yields
\begin{align}
e^{i\hat{A}\delta t} e^{i\hat{B}\delta t} e^{-i\hat{A}\delta t} e^{-i\hat{B}\delta t}
= e^{[\hat{A},\hat{B}]\delta t^2} + \mathcal{O}(\delta t^3),
\end{align}
demonstrating that, in the limit $\delta t \to 0$, we effectively realizes an evolution under the Hamiltonian $i[\hat{A},\hat{B}]$ for a duration proportional to $\delta t^2$. 

Following \cite{fukui2022building}, higher powers of $\hat{x}$ and symmetric operator products can be exactly constructed through nested commutator relations as
\begin{align}
&\hat{x}^{m+1} = -\frac{2}{3m} \Big[\hat{x}^m, \big[\hat{x}^3, \hat{p}^2\big]\Big], \\
&\hat{x}^m \hat{p}^n + \hat{p}^n \hat{x}^m =
\frac{-4i}{(n+1)(m+1)}\big[ \hat{x}^{m+1}, \hat{p}^{n+1} \big]
- \frac{1}{n+1} \sum_{k=1}^{n-1} \big[ \hat{p}^{n-k}, \big[ \hat{x}^m, \hat{p}^k \big] \big],
\end{align}
showing that arbitrary polynomial Hamiltonians can be generated through nested commutator sequences of Gaussian and non-Gaussian gates. Together, these relations demonstrate how the universal set suffices to construct any finite-degree polynomial Hamiltonian in $\hat{x}$ and $\hat{p}$, thereby enabling the approximation of arbitrary unitaries. 

The Wigner distribution $W(x,p)$ provides a complete phase-space representation of a quantum state and is defined for a pure state as  
\begin{align}
W(x,p) = \frac{1}{\pi \hbar} \int_{-\infty}^{\infty} 
\psi^\star(x+y)\psi(x-y)e^{2ipy/\hbar} \, dy,
\end{align}
where $\psi(x)$ is the wavefunction and $(x,p)$ represents the position and momentum. Unlike classical probability distributions, the Wigner function can take negative values, which serve as a clear signature of nonclassicality and quantum interference effects. Such negative regions are inherently small, bounded by the uncertainty principle, and vanish in the classical limit \cite{wigner1932quantum, bastiaans1979wigner}. Gaussian states have strictly non-negative Wigner functions that follow a Gaussian distribution in phase space and are fully characterized by the first and second moments of the quadrature operators (mean vector and covariance matrix). In contrast, non-Gaussian states exhibit Wigner functions with negative regions, reflecting their genuinely quantum nature. These states are essential for universal quantum computation and quantum error correction, as they introduce the required nonlinearity beyond Gaussian operations. Their preparation typically involves nonlinear interactions or conditional measurements, such as photon subtraction, addition, or cubic phase operations.
\subsection{Complex Phase Space Representation for CV Optimization}
In this section, we establish how complex-variable optimization naturally maps onto CV quantum systems and show how CV gate operations can be used to construct the CCV-QAOA framework.

Let $\mathbb{C}^n$ denote the space of $n$-dimensional complex vectors. Each complex number can be expressed as $z = x + i y$, where $x = \Re(z)$ and $y = \Im(z)$ are its real and imaginary components, respectively \cite{stein2010complex}. The conjugate is $\bar{z} = x - i y$, while the transpose and conjugate transpose are denoted by $z^T$ and $z^H$, respectively. The Euclidean norm is given by $\|z\| = \sqrt{z z^H} = \sqrt{x^2 + y^2}$. We consider the minimization of the unconstrained optimization problem over complex variables:
\begin{align}
    \min_{z \in \mathbb{C}^n} f(z),
    \label{pb:1}
\end{align}
where $f(z)$ is expressed using CR (Wirtinger) calculus \cite{remmert1991theory, brandwood1983complex}, exploiting the duality between $\mathbb{C}$ and $\mathbb{R}^2$. Let $z = x + i y$, we equivalently define $f(z) = f(x,y)$, with $f : \mathbb{R}^n \times \mathbb{R}^n \to \mathbb{R}$. This representation preserves the analytic structure of the complex domain while enabling optimization within a real-valued formulation.

\begin{figure}[ht]
    \centering
    \includegraphics[width=0.75\linewidth]{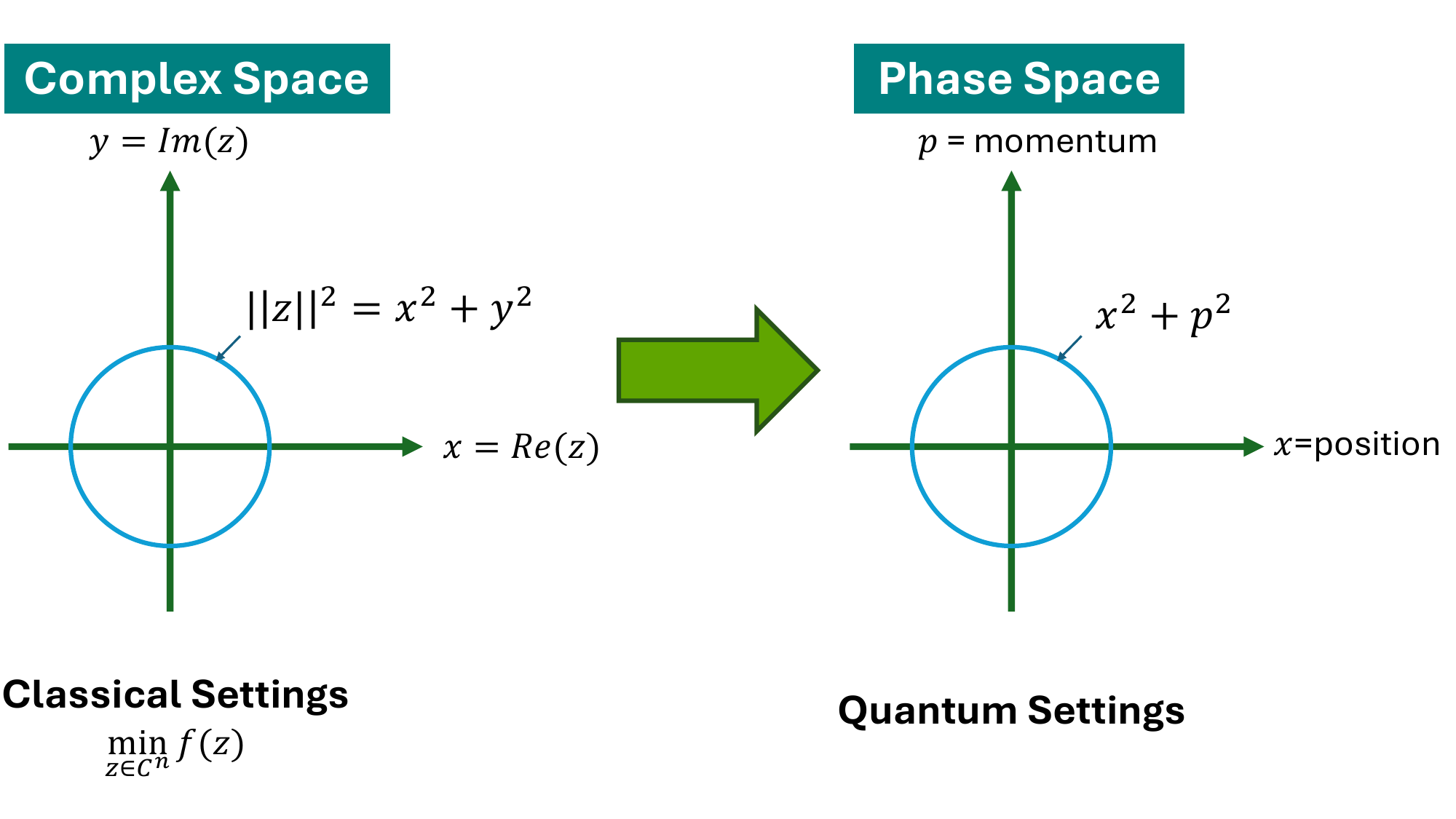}
    \caption{Mapping from complex space to phase space: the real component $\Re(z)$ corresponds to the position quadrature $\hat{x}$, and the imaginary component $\Im(z)$ to the momentum quadrature $\hat{p}$}
    \label{fig: complex space to phase space}
\end{figure}

Each variable $z_i \in \mathbb{C}$ decomposes into its real and imaginary parts, providing two continuous degrees of freedom. In the phase-space picture of quantum mechanics, these degrees of freedom map naturally to the position and momentum quadratures of a bosonic mode. Hence, $\Re(z) \leftrightarrow \hat{x}$ and $\Im(z) \leftrightarrow \hat{p}$, as illustrated in Fig.~\ref{fig: complex space to phase space}. The left diagram represents the complex plane, while the right shows the corresponding phase-space picture. For example, the simple norm $|z|^{2} = x^{2} + y^{2}$ in complex space maps directly to the quadratic form $x^{2} + p^{2}$ in phase space, highlighting the one-to-one correspondence between complex variables and CV quadratures.

The algorithm starts in the vacuum state $\ket{0}$, followed by a squeezing operation that enhances convergence. A squeezing parameter $r$ controls photon number $N$ statistics:
\begin{align}
\langle N \rangle = \sinh^2 r, \quad 
\mathrm{Var}(N) = 2\sinh^2 r\,(\sinh^2 r + 1),
\end{align}
Larger $r$ values increase the mean and variance, populating higher Fock levels \cite{walls1983squeezed}. The cost Hamiltonian $\hat{H}_C = f(\hat{x}, \hat{p})$ encodes the objective function, while a mixer Hamiltonian $\hat{H}_M$ is defined such that $[\hat{H}_C, \hat{H}_M] \neq 0$, ensuring nontrivial exploration of the search space. The corresponding variational ansatz alternates between these unitaries, as shown in equation \eqref{qaoa}. Quadrature measurements are performed to estimate the cost expectation. Heterodyne detection yields simultaneous estimates of $\hat{x}$ and $\hat{p}$ within a single phase, optimal for Gaussian states but limited by vacuum noise, whereas homodyne detection measures each quadrature separately under complementary squeezing configurations \cite{collett1987quantum}. The classical optimizer iteratively updates the variational parameters $(\boldsymbol{\gamma}, \boldsymbol{\beta})$ to minimize the expectation value of the cost Hamiltonian as in \eqref{expectation}, until convergence, yielding the optimized quantum state $\ket{\psi(\boldsymbol{\gamma}^\star, \boldsymbol{\beta}^\star)}$. We then sample from this optimised state to generate candidate solutions $z$, and the best sample, according to the classical objective, is reported as the approximate solution.

To characterize the final state $\ket{\psi(\boldsymbol{\gamma}^\star, \boldsymbol{\beta}^\star)}$, we compute its Wigner function, which serves as a characterization tool, revealing how the distribution localizes around optimal regions and whether the variational circuit accurately reproduces the structure of the target landscape. This representation is central to evaluating the accuracy and expressivity of CCV-QAOA in both Gaussian and non-Gaussian regimes.
\section{CCV-QAOA Quantum Algorithm}\label{sec3}
This section presents the complete formulation of the CCV-QAOA. Building on the CV formalism introduced in the previous section, the algorithm performs variational quantum optimization in the complex domain through parametrized evolutions of a bosonic system. 

\begin{figure}[ht]
    \centering
    \includegraphics[width=1\linewidth]{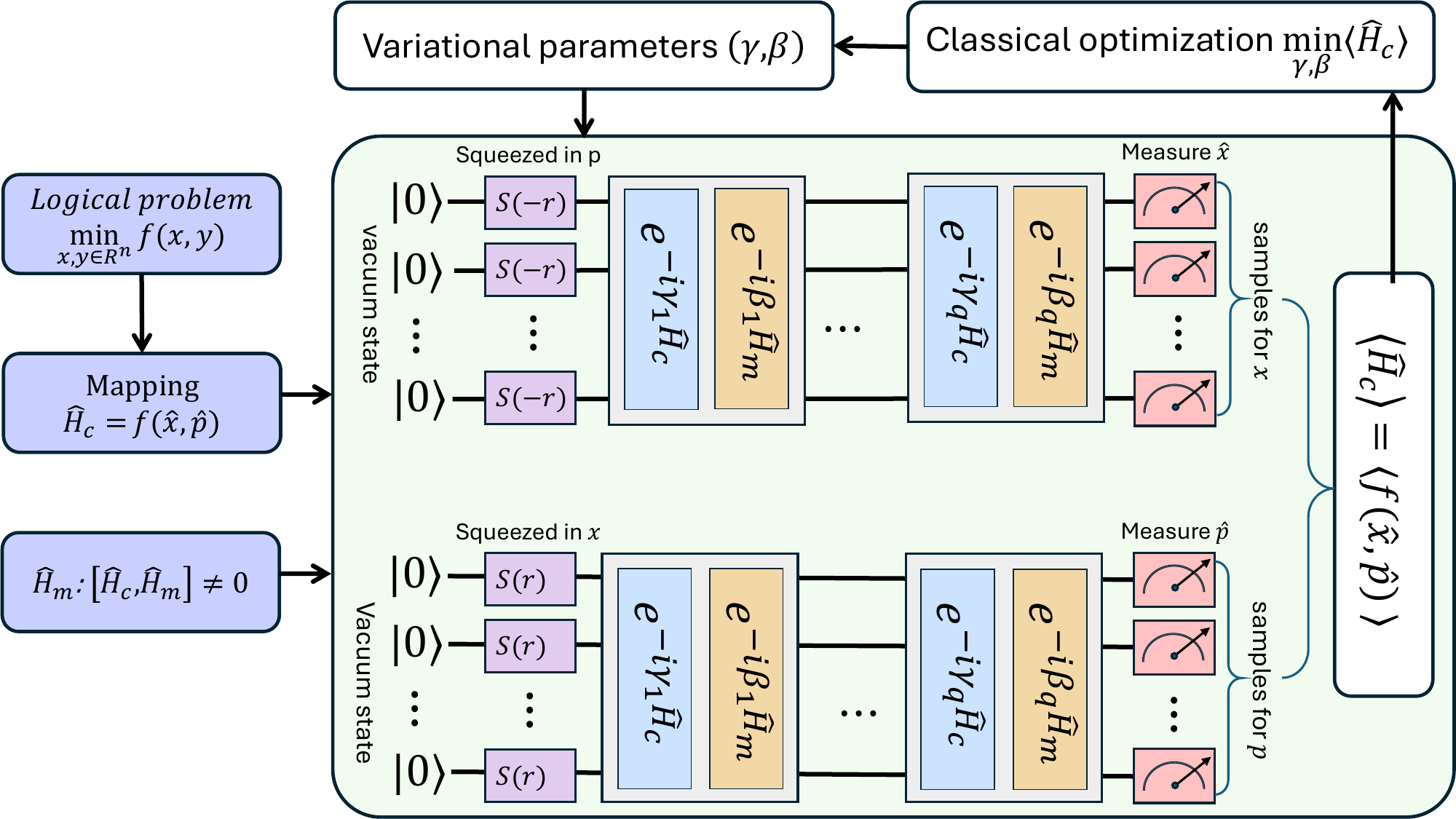}
    \caption{Workflow of CCV-QAOA: initialization, circuit evolution, measurement, and classical optimization in a feedback loop}
    \label{fig: CCV-QAOA}
\end{figure}
\begin{algorithm}[tb]
\caption{CCV-QAOA}
\label{alg:cvqaoa-algpx}
\begin{algorithmic}[1]
\Require $f(z)$; depth $q$; shots $N$; squeezing parameter $r$; backend $\mathsf{B}$; 
        cutoff $D$ if Fock; optimizer \textsf{Opt}; tolerance $\varepsilon$; max iters $T_{\max}$.
\Ensure $(\boldsymbol{\gamma}^\star,\boldsymbol{\beta}^\star)$, final samples

\State Encode cost Hamiltonian $\hat{H}_C \gets f(\hat{x},\hat{p})$
\State Choose mixer Hamiltonian $\hat{H}_M$ with $[\hat{H}_M,\hat{H}_C]\neq 0$
\State Initialize $(\boldsymbol{\gamma}^{(0)},\boldsymbol{\beta}^{(0)})$

\For{$t=0$ to $T_{\max}-1$}

  \State Build $U_j(\gamma_j^{(t)},\beta_j^{(t)}) \gets 
         e^{-i\beta_j^{(t)}\hat H_M} e^{-i\gamma_j^{(t)}\hat H_C}$ for $j=1,\cdots,q$

  \State Prepare initial squeezed vacuum state $\ket{\psi_0}$

  \State $\ket{\psi^{(t)}} \gets \Big(\prod_{j=1}^q U_j\Big)\ket{\psi_0}$

    \If{$\mathsf{B}=\text{Gaussian}$}
        \State Estimate $\widehat{\mathcal{L}}^{(t)}$ via \textbf{heterodyne} (one phase)
    \Else
        \State Estimate $\widehat{\mathcal{L}}^{(t)}$ via \textbf{homodyne} (two phases)
    \EndIf
    
  \If{$t>0$ \textbf{and} $|\widehat{\mathcal{L}}^{(t)} - \widehat{\mathcal{L}}^{(t-1)}| < \varepsilon$}
      \State \textbf{break}
  \EndIf

  \State $(\boldsymbol{\gamma}^{(t+1)},\boldsymbol{\beta}^{(t+1)}) 
         \gets \textsf{Opt}\big((\boldsymbol{\gamma}^{(t)},\boldsymbol{\beta}^{(t)}), \widehat{\mathcal{L}}^{(t)}\big)$

\EndFor

\State $(\boldsymbol{\gamma}^\star,\boldsymbol{\beta}^\star) \gets (\boldsymbol{\gamma}^{(t)},\boldsymbol{\beta}^{(t)})$
\State \textbf{return} $(\boldsymbol{\gamma}^\star,\boldsymbol{\beta}^\star)$ and final samples from 
       $\ket{\psi(\boldsymbol{\gamma}^\star,\boldsymbol{\beta}^\star)}$
\State \textbf{post-process:} compute Wigner distribution $W(x,p)$ of 
       $\ket{\psi(\boldsymbol{\gamma}^\star,\boldsymbol{\beta}^\star)}$

\end{algorithmic}
\end{algorithm}

In Figure~\ref{fig: CCV-QAOA}, we summarize the full CCV-QAOA workflow. We begin with a continuous objective function \(f(z)\), where \(z=x+iy\), and encode it into a cost Hamiltonian \(\hat{H}_C = f(\hat{x},\hat{p})\). A mixer Hamiltonian \(\hat{H}_M\), chosen to satisfy \([\hat{H}_C,\hat{H}_M]\neq 0\), defines the alternating parametrized layers of the quantum circuit together with \(\hat{H}_C\). If both quadratures \(\hat{x}\) and \(\hat{p}\) are measured simultaneously using heterodyne detection, the algorithm operates in a single measurement phase. Otherwise, CCV-QAOA proceeds in two phases. In the first phase (real-part measurement), we prepare the vacuum state \(\ket{0}\), apply momentum-squeezing (negative squeezing parameter) to expand the wavefunction along the \(x\)-axis, then apply \(q\) alternating cost and mixer layers before performing homodyne detection of \(\hat{x}\) to obtain real-valued samples. In the second phase (imaginary-part measurement), we repeat the same circuit but begin with \(x\)-squeezing (positive squeezing parameter), enhancing resolution in momentum. After the same \(q\)-layer evolution, homodyne detection of \(\hat{p}\) yields momentum samples. Combining the two sets of samples provides paired \((x,p)\) values that approximate the expectation of the cost Hamiltonian. These estimates feed into the classical optimizer during the variational loop.

Algorithm~\ref{alg:cvqaoa-algpx} sketches the iterative optimization process. To run the algorithm, the user provides the objective function $f(z)$, the circuit depth $q$, the number of shots $N$, the squeezing parameter $r$, the choice of backend $\mathsf{B}$ (Gaussian or Fock), the cutoff dimension $D$ when using the Fock backend, and a classical optimization method with stopping criteria given by the tolerance $\varepsilon$ and the maximum number of iterations $T_{\max}$. With these inputs, the algorithm constructs the cost Hamiltonian $\hat{H}_C=f(\hat{x},\hat{p})$, selects a mixer Hamiltonian $\hat{H}_M$ satisfying $[\hat{H}_C,\hat{H}_M]\neq0$, and initializes the variational parameters $(\boldsymbol{\gamma}^{(0)},\boldsymbol{\beta}^{(0)})$.

At each iteration $t$, the algorithm builds the depth-$q$ unitary blocks $U_j(\gamma_j^{(t)},\beta_j^{(t)})$, prepares the squeezed vacuum state, and applies the alternating cost–mixer layers to obtain $\ket{\psi^{(t)}}$. The cost estimate $\widehat{\mathcal{L}}^{(t)}$ is then computed from $N$ quadrature samples using the measurement rule defined by the backend: heterodyne sampling (one phase) if $\mathsf{B}$ is Gaussian, or homodyne sampling (two phases) if $\mathsf{B}$ is Fock. The resulting sampled cost is passed to the classical optimizer, which updates the parameters to $(\boldsymbol{\gamma}^{(t+1)},\boldsymbol{\beta}^{(t+1)})$. The loop repeats until the stopping criterion is reached, either the cost change falls below $\varepsilon$, or the iteration limit $T_{\max}$ is met.

Once convergence is achieved, the algorithm outputs the optimized parameters $(\boldsymbol{\gamma}^\star,\boldsymbol{\beta}^\star)$ and the final samples from $\ket{\psi(\boldsymbol{\gamma}^\star,\boldsymbol{\beta}^\star)}$. We then use this optimised state to generate samples of the decision variable $z$ by repeatedly executing the final circuit, and the best sample, according to the classical objective function, is reported as the approximate solution. The optimized quantum state is then visualized through its Wigner function $W(x,p)$, allowing for inspecting how the state localizes around the optimal regions and whether the variational circuit produces non-Gaussian or structured features relevant to the underlying landscape.

The computational backend $\mathsf{B}$ used in the implementations determines how quantum states and operations are represented \cite{killoran2019strawberry}. The Gaussian backend provides an efficient covariance-matrix description suitable for circuits composed exclusively of Gaussian gates. In contrast, the Fock backend enables universal modeling by incorporating non-Gaussian gates, at the cost of a truncated Hilbert space with cutoff dimension $D$. A smaller cutoff \(D\) reduces computational cost but increases truncation error, whereas increasing \(D\) improves accuracy at the expense of substantially greater computational resources.

The optimization of variational parameters in CCV-QAOA can be efficiently performed using the Covariance Matrix Adaptation Evolution Strategy (CMA-ES) \cite{hansen2001completely, hansen2016cma}, a stochastic, derivative-free optimization method particularly well suited for noisy and non-convex landscapes typical of quantum variational algorithms. CMA-ES iteratively samples a population of candidate parameter vectors from a multivariate Gaussian distribution, evaluates their cost function, and updates both the mean and covariance of the sampling distribution toward regions of lower cost. This adaptive mechanism enables efficient exploration of rugged energy surfaces without relying on gradient information, which can be unreliable or costly to estimate in quantum hardware.

\section{Optimization Use Cases}\label{sec4}
In this section, we present a series of optimization case studies that demonstrate how continuous-variable (CV) quantum computing, and in particular the proposed CCV-QAOA framework, can be applied to different classes of optimization problems. We begin by describing the simulation setup and the basic assumptions of the model, and then move through convex, constrained, and non-convex examples.

All simulations were performed using the \emph{Strawberry Fields} software platform \cite{killoran2019strawberry}, an open-source Python library for photonic quantum computing developed by Xanadu \cite{xanadu}. Experiments were executed on a standard personal computer running Python 3, equipped with an 11th-generation Intel Core i7-1185G7 CPU at 3.00~GHz and 32~GB of RAM.

Throughout the use cases, we employ the kinetic mixer Hamiltonian $\hat{H}_M = \hat{p}^2$, but other choices are also possible. The choice of backend depends on the circuit structure. For purely Gaussian circuits, we use the Gaussian backend with heterodyne detection, which naturally provides simultaneous estimates of $\hat{x}$ and $\hat{p}$. For circuits that include non-Gaussian gates (e.g., Kerr), we use the Fock backend with homodyne detection \cite{collett1987quantum}, measuring $\hat{x}$ and $\hat{p}$ in separate phases.

In practice, the exact expectation value of the cost Hamiltonian cannot be computed analytically and is instead approximated from a finite set of quadrature samples. Given $N$ measurement outcomes $(x^{(k)},p^{(k)})$, the empirical estimator at iteration $t$ is
\begin{align}
\widehat{\mathcal L}^{(t)} 
= \frac{1}{N} \sum_{k=1}^{N} f(x^{(k)},p^{(k)}).
\end{align}
This sampling-based estimator follows the standard paradigm in variational quantum algorithms, where expectation values are replaced by sample averages over repeated circuit executions (see, e.g.,~\cite{peruzzo2014variational, schuld2019quantum}). The variational parameters $({\gamma}, {\beta})$ are then updated using the CMA-ES algorithm.

\textbf{Assumptions for the model.}
\begin{itemize}
\item The optimization problem admits at least one global minimum.
\item The optimization problem is polynomial in $z=(x,y)$, or admits a finite-series expansion that can be implemented as a finite-degree Hamiltonian.
\end{itemize}

\subsection{Convex Unconstrained Optimization: Quadratic Problem}

We first consider the simplest setting: a convex unconstrained problem, where every local minimum is also a global minimum~\cite{bertsekas2003convex}. An example is quadratic minimization
\begin{align}
    \min_{z \in \mathbb{C}^n} \quad z^H A z + \Re(c^H z),
    \label{eq:convex}
\end{align}
with $A \succeq 0$ (positive semidefinite). We evaluate the performance of CCV-QAOA across four scenarios: (i) a single illustrative instance, (ii) scaling with circuit depth and problem size, (iii) sensitivity to the cutoff dimension, and (iv) comparison with the CV-QAOA algorithm proposed in \cite{verdon2019quantum}.

\subsubsection{Single Quadratic Instance}

For $n=3$, we select $A = I_3$ and $c=-(4+4i)\mathbf{1}_3$, where $I_3$ is the $3\times 3$ identity matrix, and $\mathbf{1}_3$ is the three-dimensional vector of ones.
The CCV-QAOA circuit is executed on the Gaussian backend with squeezing parameter $r=0.6$, depth $q=6$, and $N=50$ shots per iteration. On the classical side, CMA-ES is run for 250 iterations.

The classical optimizer yields the global minimum $z^\star=(2+2i,\,2+2i,\,2+2i)$ with classical objective value $-24$. CCV-QAOA reaches a best sampled objective of $-23.7$, with an approximate solution \[z^\star_{\text{QAOA}}=(2.1+2.2i,\,2.1+1.9i,\,1.7+1.7i)\]
and total runtime of $105$ seconds.
\begin{figure}[ht]
    \centering
    \includegraphics[width=1\linewidth]{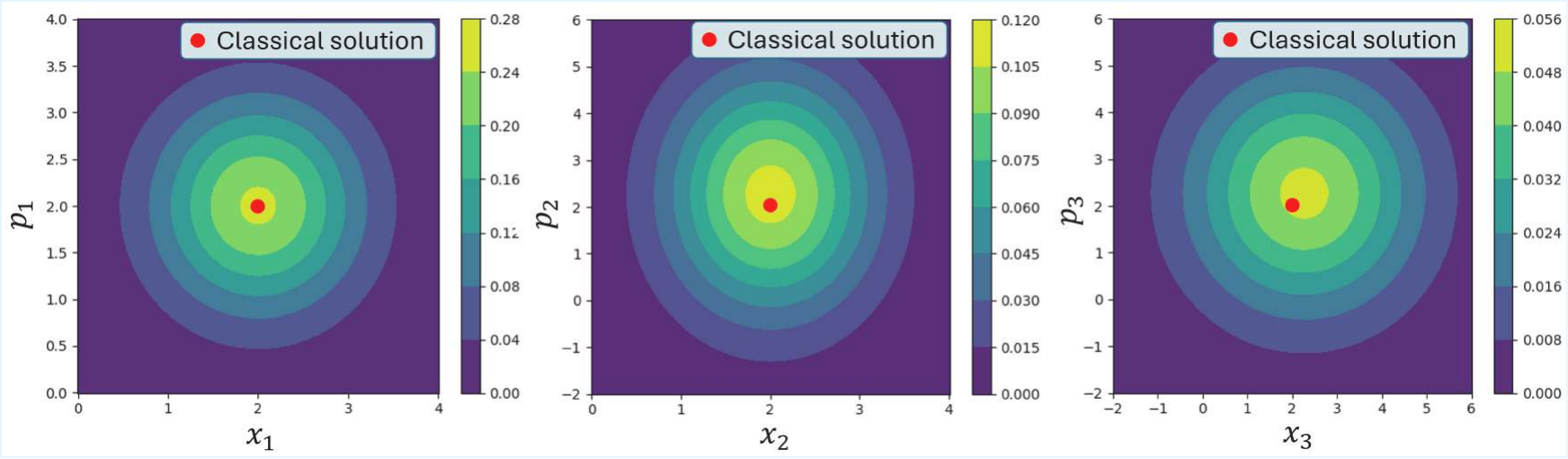}
    \caption{Final-state Wigner distributions for the quadratic instance. Left: first qumode; middle: second qumode; right: third qumode. The red point indicates the classical optimal solution}
    \label{fig:quadratic-states}
\end{figure}

\begin{figure}[ht]
    \centering
    \includegraphics[width=0.5\linewidth]{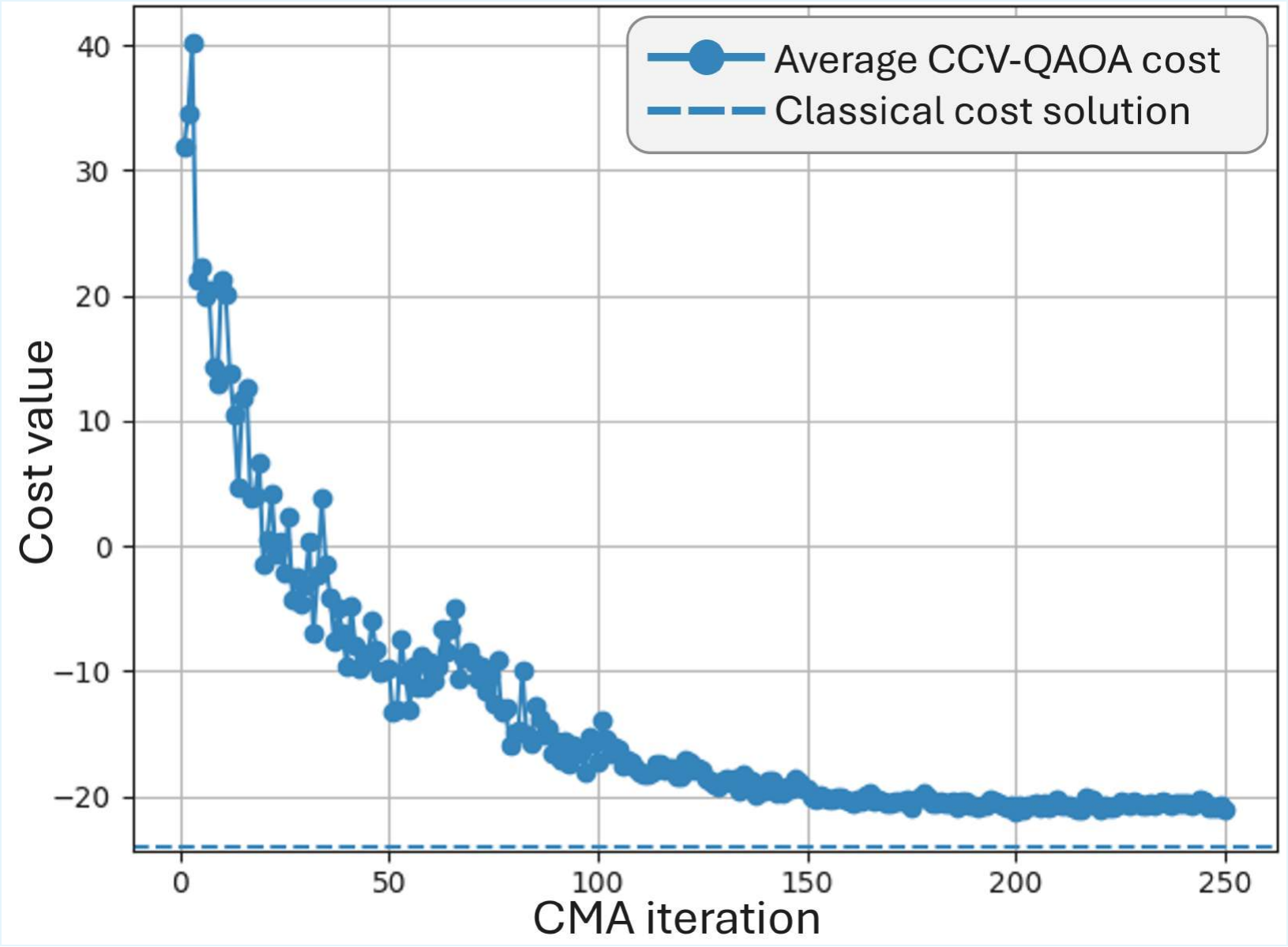}
    \caption{Convergence of the CCV-QAOA objective across CMA-ES iterations for the quadratic instance}
    \label{fig:quadratic-convergence}
\end{figure}
Figure~\ref{fig:quadratic-states} shows the final Wigner distributions for the three qumodes, illustrating how the quantum state concentrates around the classical optimum in phase space. Figure~\ref{fig:quadratic-convergence} presents the CMA-ES optimization trace, confirming stable and almost monotonic convergence of the sampled objective across iterations.

\subsubsection{Scaling with Circuit Depth and Problem Size}

We benchmark CCV-QAOA for depths \(q \in \{2,4,6\}\) and problem sizes \(n \in \{1,2,3,4\}\). Table~\ref{table 1} reports, for each configuration, the achieved cost, the final success probability, and the runtime, averaged over 10 runs. The first column lists the problem dimension \(n\), the second gives the classical optimal value, and the remaining columns (grouped by circuit depth \(q\)) provide the best cost obtained (``Cost''), the success probability (``Succ.~P''), and the total runtime in seconds (``Time'').

\begin{figure}[ht]
    \centering
    \includegraphics[width=1\linewidth]{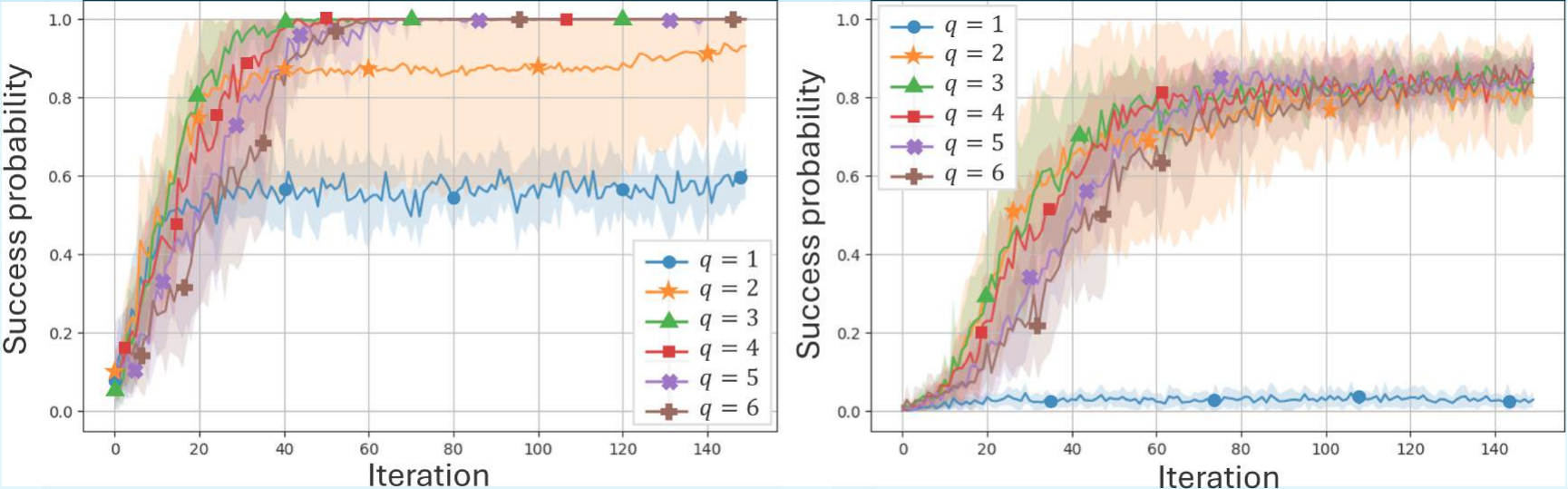}
    \caption{Success probability per iteration for different depths $q \in \{1,2,3,4,5,6\}$. 
    Left: $n=2$; right: $n=4$}
    \label{fig:success-prob}
\end{figure}

The results show that increasing the depth generally improves convergence and success probability, especially for larger problem sizes. For $n=1$ and $n=2$, even shallow circuits already achieve near-optimal performance, while for $n=3$ and $n=4$, deeper circuits help maintain high success probability and accurate cost values. Figure~\ref{fig:success-prob} further illustrates the evolution of the success probability over iterations for $n=2$ and $n=4$, confirming that deeper circuits yield more stable and robust optimization trajectories.

\begin{table}[ht]
\centering
\caption{Benchmark results for depths $q=2,4,6$: achieved cost, success probability, and runtime (averaged over 10 runs)}
\begin{tabular}{|c|c|ccc|ccc|ccc|}
\toprule
\multirow{2}{*}{$n$} & \multirow{2}{*}{Optimal} 
& \multicolumn{3}{c|}{$q=2$} 
& \multicolumn{3}{c|}{$q=4$} 
& \multicolumn{3}{c|}{$q=6$} \\
\cmidrule(lr){3-5}\cmidrule(lr){6-8}\cmidrule(lr){9-11}
 &  & Cost & Succ.~P & Time 
    & Cost & Succ.~P & Time
    & Cost & Succ.~P & Time \\
\midrule
1 & -8.00  & -7.99 & 1.00 & 7.19 
           & -8.00 & 1.00 & 16.44
           & -8.00 & 1.00 & 51.62 \\
2 & -16.00 & -15.84 & 0.93 & 10.55 
           & -15.91 & 1.00 & 21.77
           & -15.90 & 1.00 & 100.59 \\
3 & -24.00 & -23.60 & 1.00 & 14.33 
           & -23.53 & 1.00 & 27.68
           & -23.66 & 0.99 & 55.42 \\
4 & -32.00 & -30.93 & 0.80 & 16.76 
           & -31.14 & 0.88 & 33.49
           & -31.10 & 0.89 & 40.72 \\
\bottomrule
\end{tabular}
\label{table 1}
\end{table}

\subsubsection{Sensitivity to the Cutoff Dimension}

We next study the impact of the cutoff dimension $D$, which controls how many Fock states are retained in the truncated Hilbert space. Table~\ref{tab:cutoff_results} reports, for several values of $D$, the best cost achieved, the success probability, and the total runtime in seconds.

The first row (``Best cost'') shows that larger cutoffs generally yield solutions closer to the classical optimum $-8.00$, with $D=15$ achieving the best approximation. The second row (``Success $P$'') indicates how much of the probability mass is concentrated near the optimal region. While this quantity remains relatively stable for moderate cutoffs, very small or very large values of $D$ can slightly reduce the concentration. The final row (``Runtime'') highlights the computational cost of increasing $D$: simulation time grows rapidly from 383 seconds at $D=5$ to 2804 seconds at $D=15$.

Overall, these results reveal a clear trade-off between fidelity and computational efficiency. Very small cutoff values (e.g., $D=2$) run quickly but lose some accuracy, whereas very large cutoffs improve precision at the expense of substantially higher runtimes. Importantly, intermediate values such as $D=5$ or $D=7$ offer a good compromise: they provide competitive objective values and maintain reasonable success probability while keeping runtimes manageable. This suggests that CCV-QAOA remains effective even under moderate truncation, which is encouraging for practical implementations with limited computational resources.

\begin{table}[tb]
    \centering
    \caption{Performance of CCV-QAOA for different cutoff dimensions $D$}
    \begin{tabular}{|c|ccccc|}
    \hline
       \textbf{Cutoff $D$} &  2 & 5 & 7 & 10 & 15 \\
       \hline
       Best cost  & -7.76 & -7.65 & -7.92 & -7.88 & -7.98 \\
       Success $P$ & 0.85 & 0.80 & 0.55 & 0.82 & 0.78 \\
       Runtime (s) & 519 & 383 & 663 & 1331 & 2804 \\
       \hline
    \end{tabular}
    \label{tab:cutoff_results}
\end{table}

\subsubsection{Comparison with CV-QAOA}

Finally, we compare CCV-QAOA with the CV-QAOA algorithm in \cite{verdon2019quantum} for problem sizes $n\in\{1,2\}$, circuit depth $q=1$, $N=15$ measurement shots, backend $\mathcal{B}=$ Fock, and cutoff dimension $D=6$. Table~\ref{tab:ccv-cv-results} summarizes the results. The first column lists the problem size $n$, and the second column gives the classical optimal cost. For each of CCV-QAOA and CV-QAOA, the columns ``Cost'', ``Succ.~P'', and ``Time'' report the best objective value, the success probability near the optimal solution, and the total runtime in seconds, respectively.

Both methods achieve costs that are very close to the classical optimum and maintain a high success probability. However, CCV-QAOA consistently yields lower runtime: for $n=1$ it is about $40\%$ faster, and for $n=2$ it is nearly three times faster than CV-QAOA. This reduction arises from the complex-variable encoding, where each complex variable is represented by a single qumode, whereas CV-QAOA encodes real variables and thus requires two qumodes to represent a complex degree of freedom. As a result, the effective Hilbert-space dimension is reduced, leading to faster simulations without sacrificing accuracy or success probability.

\begin{table}[tb]
\centering
\caption{Comparison of CCV-QAOA and CV-QAOA at depth $q=1$: achieved cost, success probability, and runtime}
\begin{tabular}{|c|c|ccc|ccc|}
\toprule
\multirow{2}{*}{$n$} & \multirow{2}{*}{Classical Cost} 
& \multicolumn{3}{c|}{CCV-QAOA} 
& \multicolumn{3}{c|}{CV-QAOA} \\
\cmidrule(lr){3-5}\cmidrule(lr){6-8}
 &  & Cost & Succ.~P & Time 
    & Cost & Succ.~P & Time \\
\midrule
1 & -8.00  & -7.99 & 1.00 & 318 
           & -8.00 & 1.00 & 529 \\
2 & -16.00 & -15.84 & 0.87 & 638 
           & -15.95 & 1.00 & 1824 \\
\bottomrule
\end{tabular}
\label{tab:ccv-cv-results}
\end{table}

\subsection{Convex Constrained Optimization}

We now move to constrained optimization, which has the form:
\begin{equation}
    \begin{aligned}
        \min_{z \in \mathbb{C}^n} ~& f(z),\\
         \text{s.t. } & g(z) = 0, ~~ h(z) \leq 0,
    \end{aligned}
    \label{pb: cp}
\end{equation}
where $f:\mathbb{C}^n\mapsto \mathbb{R}$ is a real-valued objective function with complex decision variable $z$, $g : \mathbb{C}^n \mapsto \mathbb{C}^m$ denotes equality constraints, and $h : \mathbb{C}^n \mapsto \mathbb{R}^l$ inequality constraints.

A standard approach is to convert the constrained problem \eqref{pb: cp} into an unconstrained one by introducing penalty terms in the cost Hamiltonian~\cite{smith1997penalty}:
\begin{align}
    \min_{z \in \mathbb{C}^n}  f(z) + V_E(z) + V_I(z).\label{pb: penalty}
\end{align}
Here, the additional terms $V_E$ and $V_I$ reshape the energy landscape so that infeasible regions become energetically costly. Intuitively, equality penalties $V_E$ create deep valleys that force the wavefunction to remain close to the feasible manifold, while inequality penalties $V_I$ introduce steep barriers that discourage violating the constraints. For equality constraints,
\begin{align}
    V_E(z) = \sum_{i=1}^m \lambda  \| g_i(z) \|^2,\label{equality}
\end{align}
while inequalities may be enforced by:
\begin{itemize}
    \item Slack variable reformulation $h(z) + s^2 = 0$, where we introduce a slack variable whose square guarantees non-negativity; in this case, the constraint is reduced to an equality and \eqref{equality} applies, or
    \item Smooth rectifier functions such as Swish~\cite{ramachandran2017searching}:
    \begin{align}
        V_I(z) = \sum_{j=1}^l R\!\big(\lambda h_j(z)\big), \quad 
        R(x) = x \cdot \sigma(x),  \sigma(x) = \tfrac{1}{1+e^{-x}},
    \end{align}
     which provide an analytic approximation to the rectifier $R(x)=\lambda\max\{0,x\}$ for large $\lambda$.
\end{itemize}

As an explicit example, a constrained quadratic program can be written as
\begin{align}
    \min_{z \in \mathbb{C}^n}&  z^H A z + \Re(b^H z), 
    \\ \text{s.t. } & Cz = d,
    \label{eq:quadratic_constrained}
\end{align}
where $A$ is Hermitian, $b \in \mathbb{C}^n$, and $C$ encodes linear equality constraints. This problem is convex, since both the objective and the feasible set are convex. To solve this problem on the Gaussian backend, we employ a penalized reformulation,
\begin{align}
    \min_{z \in \mathbb{C}^n} \quad 
    z^{H} A z + \lambda \| B z - c \|^2,
\end{align}
with $\lambda > 0$ denoting the penalty parameter.

For the specific instance
\[
A =
\begin{bmatrix}
1 & 0.5 - \mathrm{i} \\
0.5 + \mathrm{i} & 2
\end{bmatrix}, 
\quad 
B = \tfrac{1}{\sqrt{10}} I_{2}, 
\quad 
c=
\begin{bmatrix}
\dfrac{1}{2\sqrt{10}} - \dfrac{\sqrt{10}}{4}\,\mathrm{i} \\[1ex]
\dfrac{3}{2\sqrt{10}} + \dfrac{1}{\sqrt{10}}\,\mathrm{i}
\end{bmatrix},
\]
the optimization was carried out with parameters depth $q=6$, squeezing $r=0.2$, backend = ``gaussian'', shots $=50$, and $300$ CMA-ES iterations. The procedure yielded an optimal cost of $-23.621$ with corresponding solution
\[
z = [\,0.57 - 11.78\mathrm{i}, \,-5.49 + 2.11\mathrm{i}\,].
\]
When sampling the optimized circuit, the approximate solution
\[
z = [\,0.475 - 12.556\mathrm{i}, \,-6.121 + 2.125\mathrm{i}\,]
\]
was obtained, with an associated cost of $-23.227$ and a total runtime of $341.95$ seconds.

\begin{figure}[ht]
    \centering
    \includegraphics[width=1\linewidth]{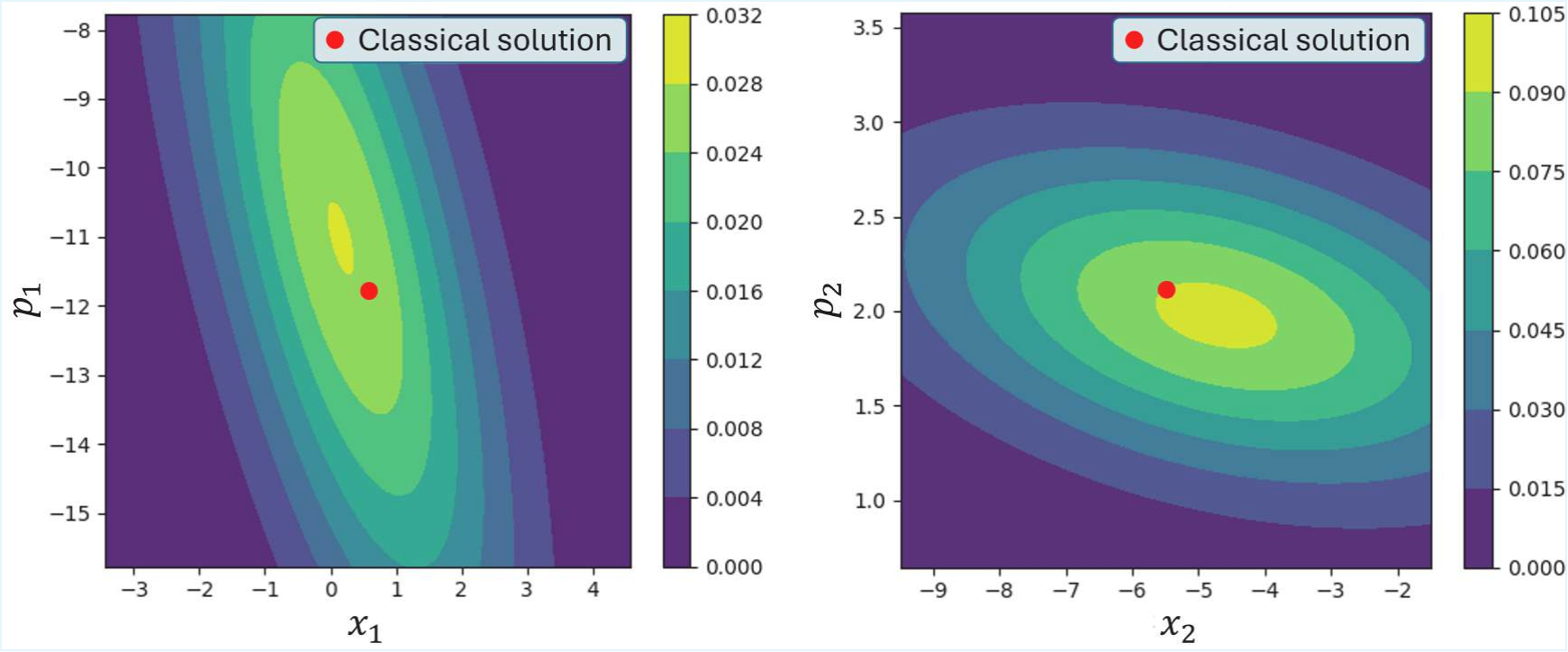}
    \caption{Constrained quadratic problem solved on the Gaussian backend. Left: 3D and 2D Wigner distributions of the first qumode. Right: 3D and 2D Wigner distributions of the second qumode. The red point is the optimal solution using the classical optimization method}
    \label{fig:constrained}
\end{figure}

\subsection{Real Multivariate Problem: Styblinski-Tang Function}
Our method also applies to multivariate functions $f(x)$ with $x \in \mathbb{R}^n$. Without loss of generality, assuming $n$ is even, we pair real variables into canonical coordinates $(x,y) \in \mathbb{R}^n$, reducing the problem to $n/2$ qumodes with quadratures $(\hat{x},\hat{p})$.

We consider the problem of minimizing the non-convex. Non-convex problems pose additional difficulties due to the presence of multiple local minima and a distinct global minimum. A local minimum is a point with no better neighboring solutions, but it may not be globally optimal. Classical optimization methods often risk getting trapped in such local minima, particularly in high-dimensional landscapes.

A canonical benchmark is the Styblinski-Tang function~\cite{styblinski1990experiments}, defined in real variables as
\begin{align}
    f(x) = \frac{1}{2}\sum_{i=1}^n \left(x_i^4 - 16x_i^2 + 5x_i\right).
\end{align}
It is highly multi-modal, with many local minima, making it challenging for gradient-based methods.

We consider the problem of minimizing the non-convex Styblinski-Tang function with $n=2$. The classical optimal solution is $f(x^\star) = -78.332$ at $x^\star=[-2.9,-2.9]$. We use the universal CV gate set to approximate the quartic gate $e^{i x^4}$ in the cost Hamiltonian.

\begin{figure}[ht]
    \centering
    \includegraphics[width=1\linewidth]{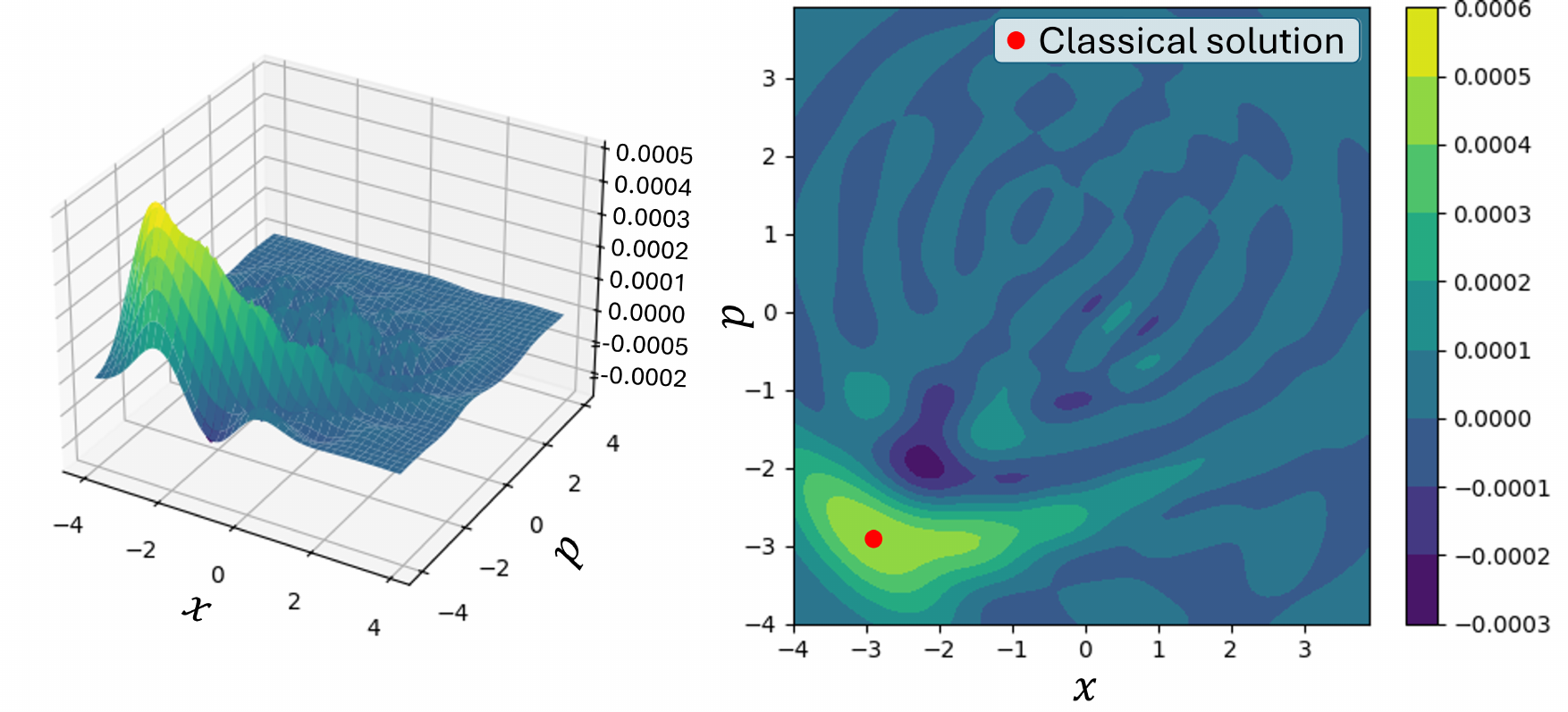}
    \caption{Wigner distribution of the final CCV-QAOA state for the real-variable Styblinski-Tang function: 3D representation (left) and 2D heatmap (right)}
    \label{fig:placeholder}
\end{figure}
For this instance, we use 100 iterations for the CMA-ES, squeezing $r=0.3$, cutoff dimension $D=14$, and $N=15$ shots. The optimal CCV-QAOA value obtained is $-78.323$, with approximate solution
\[
z_{\text{QAOA}}=[-2.914-2.883i],
\]
which is very close to the classical optimum, as shown in Figure~\ref{fig:placeholder} visualizes the final CCV-QAOA output state through its 3D and 2D Wigner representations.

\subsection{Complex Non-Convex Problem: Quartic Function}

In the complex-variable setting, we adapt this idea with the following test function:
\begin{align}
    \min_{z\in\mathbb{C}} \quad \|z-\hbar\|^{4}-\|bz\|^{2}+\Re(cz),
    \label{eq:nonconvex}
\end{align}
which preserves the quartic nonlinearity and multiple local minima, and thus serves as a complex non-convex benchmark for CCV-QAOA. Here $\hbar,c\in\mathbb{C}$ are fixed constants and $|\cdot|$ denotes the complex modulus. The quartic term creates multiple wells, while the quadratic and linear terms shift and tilt the landscape, yielding a nontrivial optimization task.

\begin{figure}[H]
    \centering
    \includegraphics[width=1\linewidth]{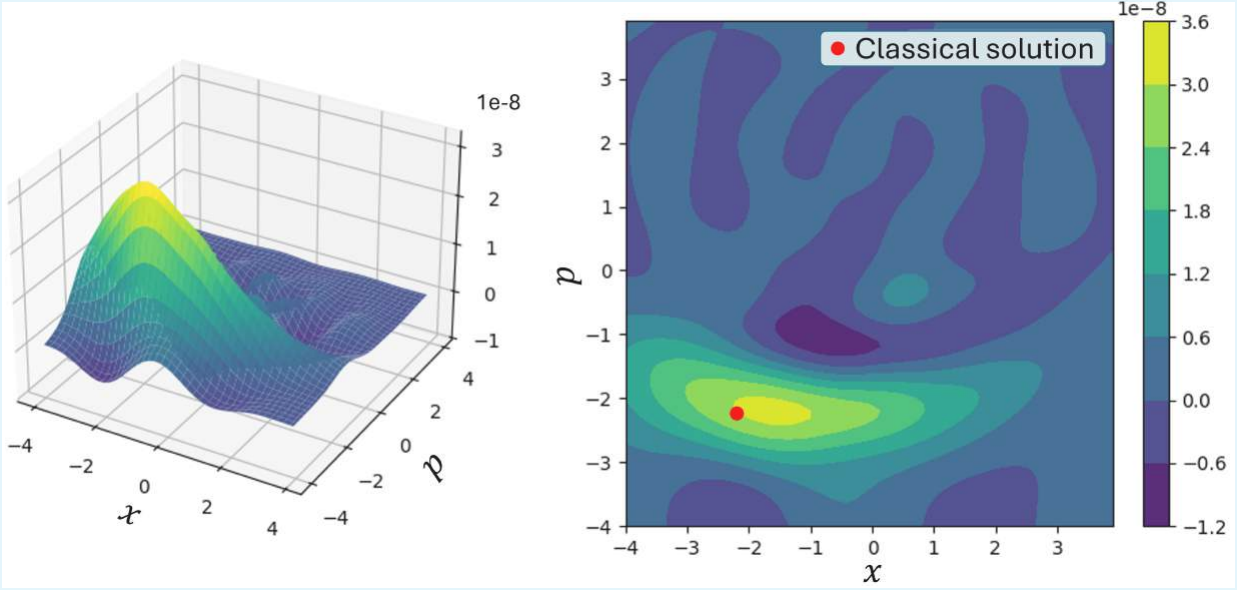}
    \caption{3D and 2D Wigner distribution of the final CCV-QAOA state for \eqref{eq:nonconvex}. The red marker denotes the classical optimizer}
    \label{fig:nonconvex}
\end{figure}

We use the Fock backend (with a Kerr gate included in the circuit), depth $q=2$, squeezing parameter $r=1$, cutoff dimension $D=10$, $N=20$ shots per iteration, and CMA-ES for 30 iterations. A classical solver returns an optimal cost of
\[
\text{Optimal cost} = -28.6969
\]
at $z^\star \approx -2.22 - 2.22\,i$. Sampling from the optimized CCV-QAOA circuit yields
\[
\text{Best sampled cost} = -28.6963,
\quad
z^\star_{\text{QAOA}} \approx -2.23 - 2.23\,i,
\]
showing excellent agreement with the classical optimum, as a final illustration, Figure~\ref{fig:nonconvex} visualizes the CCV-QAOA output state, 
showing how its Wigner distribution concentrates near the classical global minimizer.

In both the real and complex-variable non-convex cases, the reconstructed Wigner functions exhibit pronounced regions of negativity, indicating the generation of highly nonclassical states within the Fock representation. These negative regions are signatures of genuine quantum coherence and interference, and their presence in the optimized states confirms that CCV-QAOA is capable of exploring and encoding complicated energy landscapes that are difficult for purely classical methods to navigate.
\section{Discussion and Conclusion}\label{sec5}
In this paper, we extended the Continuous-Variable Quantum Approximate Optimization Algorithm (CV-QAOA) \cite{verdon2019quantum} by adapting the cost Hamiltonian to include both position $\hat{x}$ and momentum $\hat{p}$ operators. This formulation allows the algorithm to natively handle functions of complex variables and multivariate real functions. The important advantage of this approach is that each complex variable is represented directly in phase space by a pair of quadratures $(\hat{x},\hat{p})$, meaning that $n$ complex variables require only $n$ qumodes rather than $2n$, and $n$ real variables require only $n/2$ qumodes. This effectively halves the mode count compared to real-variable encodings, reducing computational scaling from $\mathcal{O}(D^N)$ to $\mathcal{O}(D^{N/2})$. Such compression is particularly relevant for non-convex continuous optimization problems.

Beyond the theoretical advantages, the optimization examples show clearly how well the method works in practice. In the convex quadratic case, CCV-QAOA always came close to the classical optimum and kept a high success rate even with small circuit depth. The scaling tests showed stable performance as the problem size and depth increased, with deeper circuits giving better results. The cutoff studies revealed a normal trade-off between accuracy and cost, but also showed that good solutions are still possible with small $D$. For constrained problems, the penalty method produced solutions that satisfied the constraints, and the Wigner plots matched the classical results closely. In non-convex problems, such as the Styblinski–Tang function and complex quartic examples, the algorithm handled multiple minima well, avoided getting stuck, and produced clear nonclassical Wigner shapes that reflected the correct global structure.

Currently, the methods focus on objective functions expressed as finite polynomials or truncated series. This restriction arises from the capabilities of existing quantum software libraries (e.g., \emph{Strawberry Fields}), which currently support only a finite set of Gaussian and non-Gaussian gates. Non-Gaussian operations such as the cubic phase gate remain highly sensitive to cutoff truncations, while the Kerr gate provides a more stable but distinct alternative. Truncating the Fock Hilbert space to $D$ states per mode introduces a trade-off: small cutoff values $D$ neglect relevant photon populations, whereas larger $D$ rapidly increase computational cost since the total dimension grows as $\mathcal{O}(D^N)$.

Increasing the circuit depth $q$ typically improves solution quality by enhancing expressivity, but also requires deeper circuits and larger squeezing parameters $r$. Higher squeezing increases the mean photon number, which in turn demands a larger cutoff dimension $D$ to faithfully capture the quantum state. However, the simulations demonstrate that the proposed algorithm maintains good accuracy and convergence even with relatively small finite cutoff values, confirming its robustness under practical computational constraints.

Overall, CCV-QAOA extends variational quantum algorithms to the complex domain, offering resource savings, natural constraint handling, and compatibility with both Gaussian and non-Gaussian gate architectures. While practical deployment will require advances in CV hardware and improved simulation efficiency, the framework establishes a systematic foundation for addressing complex-valued optimization problems with future quantum technologies.

\bibliographystyle{plainnat}
\bibliography{bib}

\end{document}